\documentclass[prd,nofootinbib,showpacs]{revtex4}
\usepackage{graphics,epsfig,epsf,psfrag}
\usepackage{amsmath,amssymb,mathrsfs,latexsym}
\usepackage{slashed}
\usepackage{bbm}
\usepackage[usenames]{color}
\usepackage{textcomp,wasysym}
\usepackage{epstopdf}

\newcommand{\y}{Y(4260)}
\newcommand{\z}{Z_c(3900)}

\begin{document}

\title{$\y$ as the first $S$-wave  open charm vector molecular state?}

\author{Martin Cleven$^{1}$\thanks{{\it Email address:} m.cleven@fz-juelich.de},
        Qian Wang$^{1}$\thanks{{\it Email address:} q.wang@fz-juelich.de},
        Feng-Kun Guo$^{2}$\thanks{{\it Email address:} fkguo@hiskp.uni-bonn.de},
        Christoph Hanhart$^{1,3}$\thanks{{\it Email address:}
        c.hanhart@fz-juelich.de}, \\
        Ulf-G. Mei{\ss}ner$^{1,2,3}$\thanks{{\it Email address:} meissner@hiskp.uni-bonn.de},
        and Qiang Zhao$^{4}$\thanks{{\it Email address:}
        zhaoq@ihep.ac.cn}\\[3mm]
  {\small
       $^1$\it Institut f\"{u}r Kernphysik and J\"ulich Center for Hadron
          Physics},\\
          {\small \it Forschungszentrum J\"{u}lich, D--52425 J\"{u}lich, Germany}\\
       {\small $^2$\it Helmholtz-Institut f\"ur Strahlen- und Kernphysik and
          Bethe Center for Theoretical Physics,}\\
       {\small \it  Universit\"at Bonn, D--53115 Bonn, Germany}\\
       {\small$^3$\it Institute for Advanced Simulation,
          Forschungszentrum J\"{u}lich, D--52425 J\"{u}lich, Germany}\\
       {\small $^4$\it Institute of High Energy Physics and Theoretical Physics Center for Science Facilities,}\\
       {\small \it Chinese Academy of Sciences, Beijing 100049, China}
          }
\date{\today}

\begin{abstract}
Since its first observation in 2005, the vector charmonium $Y(4260)$ has turned
out to be one of the prime candidates for an exotic state
in the charmonium spectrum.
It was recently proposed that the $Y(4260)$ should have a prominent
$D_1\bar{D}+c.c.$ molecular component that is strongly correlated with the
production of the charged $Z_c(3900)$.  In this paper we demonstrate that the
nontrivial cross section line shapes of $e^+e^-\to J/\psi\pi\pi$ and $h_c\pi\pi$
can be naturally explained by the molecular scenario. As a consequence we find a significantly
smaller mass for the $\y$ than previous studied. In the $e^+e^-\to \bar D D^*\pi+c.c.$ process, 
with the inclusion of an additional $S$-wave $\bar D^*\pi$ contribution constrained
 from data on the $D\bar D^*$ invariant mass distribution,
we obtain a good agreement with the data of the angular distributions.
We also predict  an unusual line shape of $\y$ in this channel that may serve as a smoking gun for 
a predominantly molecular nature of $\y$. 
Improved measurements of these observables are therefore crucial for a better understanding
of the structure of this famous resonance.
\end{abstract}

\maketitle
\section{introduction}\label{sec:introduction}

Ever since its discovery in 2005~\cite{Aubert:2005rm}, the nature of the vector
charmonium state $\y$ has remained mysterious~\cite{Brambilla:2010cs}. Most
recently the BESIII Collaboration reported a stunning result---the observation 
of a charged
charmonium state $Z_c(3900)$ in the invariant mass spectrum of $J/\psi\pi^\pm$
in $e^+e^-\to \y\to J/\psi\pi^+\pi^-$~\cite{Ablikim:2013mio}. The same 
discovery was also made by the Belle Collaboration~\cite{Liu:2013dau} and soon 
confirmed by an analysis of the CLEO-c data~\cite{Xiao:2013iha}. The $Z_c(3900)$ 
became the first
confirmed flavor exotic state in the heavy quarkonium mass region. The
continuous study of other channels, i.e. the $(D^{*} \bar{D}^{*})^{\pm}
\pi^\mp$~\cite{Ablikim:2013emm} and $e^+e^-\to
h_c\pi\pi$~\cite{Ablikim:2013wzq}, turned out to be rewarding, since
evidence for additional charged charmonium states $Z_c(4020/4025)$ was observed.

There is no doubt that the experimental observations of the $Z_c(3900)$ and
$Z_c(4020/4025)$ provide a great opportunity for understanding the strong
interaction dynamics which accounts for the formation of meson states beyond the
simple $q\bar{q}$ constituent picture. Meanwhile, it has also been recognized
that the formation of the charged $Z_c(3900)$ could shed light on the nature of
the $Y(4260)$. In Ref.~\cite{Wang:2013cya} the authors have discussed the
possibility that the $Y(4260)$ is a $D_1(2420)\bar{D}$
\footnote{Its charged conjugate part is implied and considered in the calculation. The 
same convention is used in the following for the other cases.} bound state while
the $Z_c(3900)$ is a $D\bar{D}^*$ molecule.

There exist many different interpretations for the
$Y(4260)$ in the literature. Among those there are proposals for its being a
hybrid~\cite{Zhu:2005hp,Kou:2005gt,Close:2005iz}, 
hadro-charmonium~\cite{Voloshin:2007dx,Dubynskiy:2008mq,Li:2013ssa},
$D_1\bar{D}$ molecule in potential
models~\cite{Ding:2008gr,Li:2013bca},  $\chi_{c0}\omega$ molecule
state~\cite{Dai:2012pb}, the $4S$
charmonium~\cite{LlanesEstrada:2005hz}, and a $J/\psi K\bar K$ 
bound system~\cite{MartinezTorres:2009xb}, etc.
It is therefore necessary to
identify observables which are sensitive to the structure of the $Y(4260)$.

A crucial point that makes the $Y(4260)$ special is that its mass is only about
a few tens of MeV below the first $S$-wave open charm threshold 
$D_1\bar{D}$~\cite{Beringer:1900zz}. Although the production of the relative 
$S$-wave $D_1\bar{D}$ pair is suppressed in the heavy quark 
limit~\cite{Li:2013yka}, there is evidence that in the charmonium mass region 
the heavy quark spin symmetry breaking could be large enough to allow  for the 
production of the $Y(4260)$ as a prominent $D_1\bar{D}$ 
molecule~\cite{Wang:2013kra}.

If the dominant component of the $Y(4260)$ wave function is indeed the $D_1\bar 
D$, non-trivial predictions should then follow. For instance, if the $Z_c(3900)$ and 
$X(3872)$ are isovector and isoscalar $D\bar{D}^*$ molecular states, 
respectively~\cite{Guo:2013sya}, the production mechanisms for the $Y(4260)\to 
Z_c(3900)\pi$ and $Y(4260)\to \gamma X(3872)$ should be closely related, which 
leads to the prediction of a large radiative decay rate for the $Y(4260)\to 
\gamma X(3872)$~\cite{Guo:2013zbw}. This result was confirmed by the most recent 
BESIII measurement~\cite{Yuan:2013lma}.

\begin{figure}[t]
\centering
\includegraphics[width=0.7\linewidth]{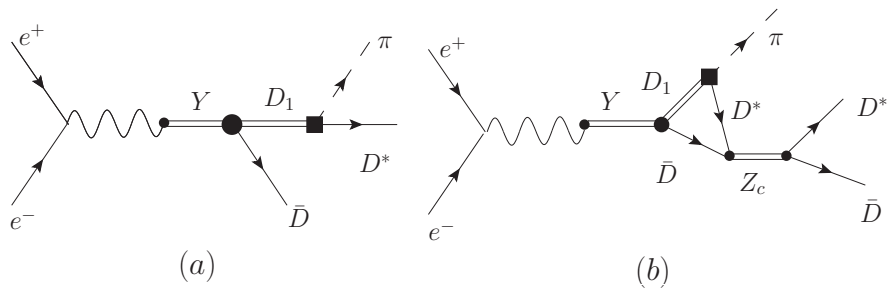}
\caption{Feynman diagrams for the $e^+e^-\to Y(4260)\to D\bar{D}^*\pi$ via
(a) the tree diagram and (b) an intermediate $Z_c$.}
\label{fig:DecayDiagrams}
\end{figure}

In this work we analyze the exclusive processes $e^+e^-\to
J/\psi\pi\pi$, $h_c\pi\pi$ and $D\bar{D}^*\pi$, in the vicinity of the
$Y(4260)$ mass region in the framework of a nonrelativistic effective field
theory~\cite{Guo:2009wr,Guo:2010ak}\footnote{For a detailed discussion of a similar 
effective field theory we refer to Ref.~\cite{mehen} and references therein.}. By treating the $Y(4260)$ as 
dominantly a
$D_1\bar{D}$  molecular state, the cross section for the $P$-wave transition
$e^+e^-\to Y(4260)\to h_c\pi\pi$ gets enhanced via intermediate meson loops
and becomes compatible with the $S$-wave transition $e^+e^-\to Y(4260)\to
J/\psi\pi\pi$~\cite{Wang:2013kra}. The $e^+e^-\to Y(4260)\to D\bar{D}^*\pi$
cross section should also be sizeable, since the $Y(4260)$ can directly
couple to  $D\bar{D}^*\pi+c.c.$ via tree diagrams, $cf$. Fig.~\ref{fig:DecayDiagrams} (a). 

The paper is structured as follows: The details of our framework are given in Sec.~\ref{sec:framework},
 numerical results and a detailed discussion follow in Sec.~\ref{sec:results}. The last section
 contains summary and outlook.

\section{theoretical framework}\label{sec:framework}

The coupling of $Y(4260)$ to $D_1\bar{D}$ in an $S$-wave are described
by the Lagrangian~\cite{Wang:2013cya,Guo:2013zbw}
\begin{eqnarray}\nonumber
\mathcal{L}_Y&=&\frac{y}{\sqrt{2}}Y^i\left(\bar{D}_a^\dag
D_{1a}^{i\dag}-\bar{D}_{1a}^{i\dag} D_a^\dag\right)+{\rm H.c.} ,
\label{eq:LagY}
\end{eqnarray}
where the (renormalized) effective coupling constant $y$ is in principle related to the probabilities of
finding the components inside the $Y(4260)$~\cite{Weinberg:1963,Baru:2003,Guo:2013zbw}, although for
the large binding energy of the $Y(4260)$ a quantitative extraction of this quantity
is hindered by the uncertainties of the method. We will
also take into account the $Z_c(3900)$ contribution the same way as
Ref.~\cite{Wang:2013cya} with $I^G(J^{PC})=1^+(1^{+-})$. The $S$-wave coupling
of the $Z_c$ to the $D\bar D^*$ is described by a Lagrangian similar to
Eq.~\eqref{eq:LagY}~\cite{Wu:2013onz},
\begin{eqnarray}\label{eq:3}
\mathcal{L}_{Z}= z ( \bar{D}_b^{*\dag i} Z_{ba}^{i} D_a^{\dag} - \bar{D}_b^{\dag} Z_{ba}^{i} D_a^{*\dag i} )+h.c.
\end{eqnarray}
with the isotriplet 
\begin{equation}\label{eq:4}
Z_{ba}=\left( \begin{array}{cc}
            \frac{1}{\sqrt{2}} Z^{0} & Z^{+} \\
            Z^{-}       &  -\frac{1}{\sqrt{2}} Z^{0}
            \end{array}  \right)_{ba}
\ ,
\end{equation}
A discussion of the bottom analogue can be found in
Refs.~\cite{Cleven:2013sq,mehen}.
So far, we can only obtain information of this coupling constant from an
analysis of the $Y(4260)\to J/\psi\pi\pi$ as performed in
Ref.~\cite{Wang:2013cya}. With $|z|\approx0.77~\mathrm{GeV}^{-1/2}$,
the branching ratio $\mathcal{B}(Z_c\to D\bar D^*)$ of about $20\%$ is
compatible with the data.

The Lagrangian describing the interaction among the $\frac 3 2^+$ and 
$\frac 12^-$ spin multiplets and pions reads~\cite{Colangelo:2005gb}
\begin{eqnarray}\nonumber
 \mathcal{L}_{D_1}&=&i\frac{h^\prime}{f_\pi}
\big[3D_{1a}^i(\partial^i\partial^j\phi_{ab})D^{*
j\dag}_b-D_{1a}^i(\partial^{j}\partial^j\phi_{ab})D_b^{* i\dag}+ ... \big]+{\rm H.c.} ,
\label{D1Dpi}\end{eqnarray}
where the coupling constant $|h'|=(0.62\pm 0.08)~\mathrm{GeV}^{-1}$ is
determined by the decay width $\Gamma_{D_1^+}=(25\pm
6)~\mathrm{MeV}$~\cite{Beringer:1900zz}. The dots indicate terms not in the focus 
of this work.

For the $\z$ we use the following propagator~\cite{Cleven:2011gp}
\begin{equation}\label{eq:propZ}
 G_Z(E)=\frac12\frac{i}{E-m_Z-\Sigma_{DD^*}(E)+i\,\tilde\Gamma_Z/2}
\end{equation}
with
\begin{eqnarray*}
 \Sigma_{DD^*}(E)
 = z^2\frac{\mu_{DD^*}^{3/2}}{8\pi}\left[\sqrt{-2\epsilon}\,
\theta(-\epsilon) - \sqrt{2\epsilon} \,
\theta(\epsilon)\right],
\end{eqnarray*}
where $\epsilon=E-m_D-m_{D^*}$ and the constant $\tilde\Gamma_Z$  accounts for
the width from decay channels other than the $D\bar D^*$ such that
the
sum of Im$(2\Sigma_Z(E))$ evaluated at the pole and
$\tilde\Gamma_Z$ gives the total
width of the $Z_c$

In the $D_1\bar D$ molecular scenario,
the propagator of $\y$ can be calculated analogous to Eq.~(\ref{eq:propZ}) with  $ \Sigma_{DD^*}(E)$ replaced by
\begin{eqnarray}
\hat \Sigma_Y\left(E\right)&=&\Sigma_{D_1D}\left(E\right)
-\mathrm{Re}\big[ \Sigma_{D_1D}\left(M_Y\right)+\left(E-M_Y\right)\partial_{E}\Sigma_{D_1D}(E)|_{E=M_Y} \big] .
\end{eqnarray}
 Here we also have two parameters, i.e. one mass $M_Y$ and one constant
residual width $\tilde\Gamma_Y$.

\section{Results}\label{sec:results}
\subsection{Line shapes of the  $\y$ in the $J/\psi\pi\pi$ and $h_c\pi\pi$ channels}

\begin{figure}[]
\hspace{40.0cm}\includegraphics[width=1\linewidth]{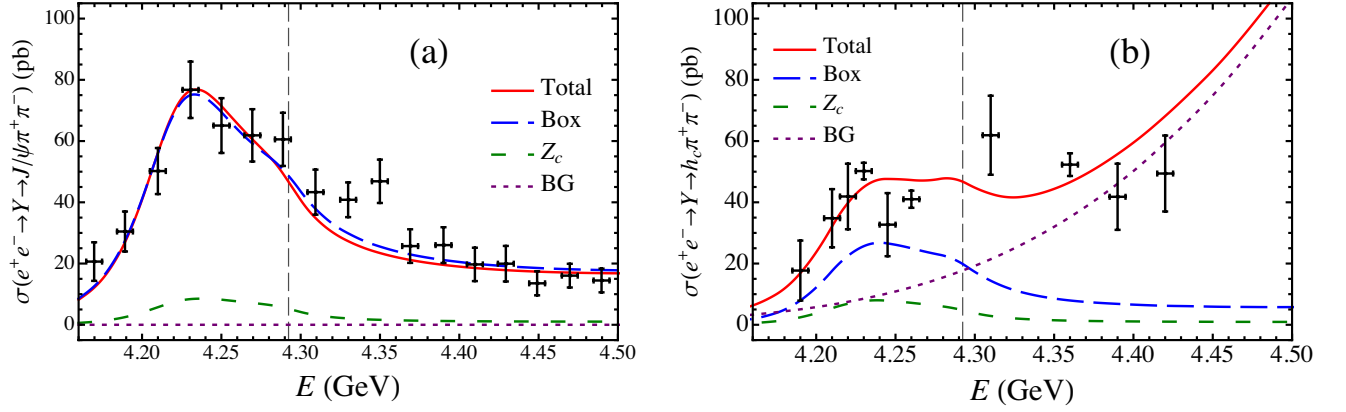}
\caption{The cross sections for the $e^+e^-\to J/\psi\pi^+\pi^-$ and
$e^+e^-\to h_c\pi^+\pi^-$ around the $Y(4260)$ mass region. The long-dashed, short-dashed, dotted and solid lines are 
the contributions from the $D_1\bar D$ box diagrams, the $\z$ pole, the $S$-wave background and the sum of them, respectively. 
The data in (a) are from Belle~\cite{Yuan:2007sj} and those
in (b) from BESIII~\cite{Ablikim:2013wzq}, respectively.}
\label{fig:JpsiHc}
\end{figure}

In this section, we present the fit results for line shapes of $e^+e^-\to Y(4260)\to J/\psi\pi\pi$ and $h_c\pi\pi$ in Fig.~\ref{fig:JpsiHc}. 
The formalism used here is a straightforward extension  of Ref.~\cite{Wang:2013cya}, i.e. including both box diagrams and
$Z_c(3900)$ pole contributions simultaneously.  It turns out that we need $S$-wave and $P$-wave background terms
for the $J/\psi\pi\pi$ and $h_c\pi\pi$ channel, respectively, in order to fit the experimental data in the energy range of
$[4.16,4.50]$~GeV for both processes. Since the partial waves of the $\pi\pi$ 
are
$S$-wave for the background and mostly $D$-wave for the box diagrams and the 
$Z_c(3900)$ pole (since the $D_1$ decays to $D\pi$ in a $D$--wave, $cf$. Eq.~(\ref{D1Dpi})),
there is no interference between them after the angular integration.
Fit parameters are the mass of the $\y$, the non-$(D_1\bar D\to D^* \bar
D\pi)$ width $\tilde\Gamma_Y$, a normalization constant and a factor for the
strength of the background in each channel. 
The combined fit gives a reduced chi-square of 2 which seems sufficient given the
simplified model used. As can be seen from 
Fig.~\ref{fig:JpsiHc} in both channels the proximity of the
 $D_1\bar D$ threshold in combination with a sizeable  $D_1\bar D \y$ coupling 
constant $y$, which is the signature
 for a dominant molecular component of $\y$, leads to an asymmetric line shape. 
Due to this asymmetry especially   in the $J/\psi\pi\pi$ channel, 
where the data are significantly better, the fit now gives a mass for the $\y$ 
significantly smaller than previous analyses, namely
\begin{equation}
M_\y = (4217.2\pm 2.0) ~\mbox{MeV}
\end{equation}
with the value of $\tilde\Gamma_Y=(55.91\pm 2.61)$ MeV, we see that the branching 
ratio for $Y(4260)\to D\bar D^*\pi $ via the intermediate $D\bar D_1$ is 
dominant within our model and can be as large as $60\%$. 
This finding is an important consistency check of our approach. Figure~\ref{fig:JpsiHc}(a) and (b) show that the
contribution of the $Z_c(3900)$ pole (short-dashed lines) is much smaller than
that of the box diagrams (long-dashed lines) which is consistent with the
results of Ref.~\cite{Wang:2013cya}.

One might question if it is sensible that the data in the $h_c\pi\pi$ channel above $4.35$ GeV
 is dominated by the background contribution. We therefore performed a series of additional 
 fits including the higher thresholds ($D_2\bar D^*$, $D_1\bar D^*$) as well as the
 $Y(4360)$, as proposed in Refs.~\cite{Aubert:2006ge,Li:2013ssa}. As expected these
 fits allowed us to remove the background contributions, however, the current data did not
 allow us to constrain sufficiently the values of the parameters. Especially with the current
 data it was not possible to decide whether a second resonance is needed. What is very important to
 this work is that regardless what dynamical content was used to fit the higher energies,
 the parameters of the $\y$ stayed largely unchanged. 
 Due to this consideration we restrict ourselves only to a detailed discussion on the simplest fit in the rest of this work.  .

\begin{figure*}
\centering \includegraphics[width=1.\linewidth]{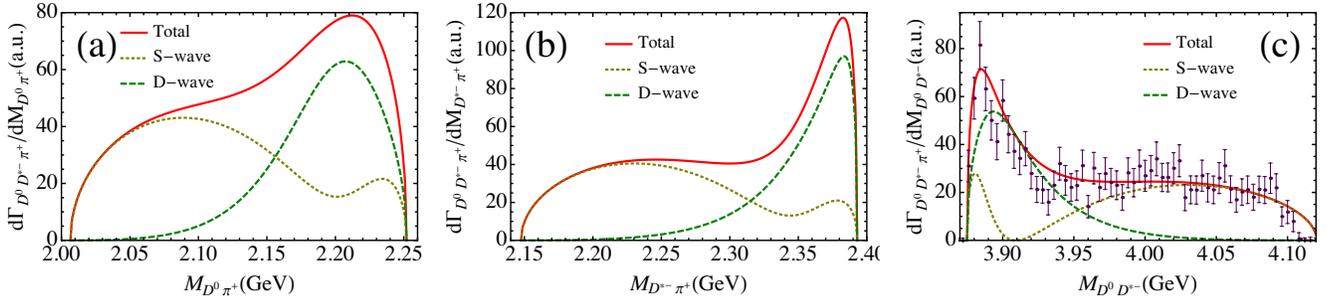}
\caption{The $D^0\pi^+$, $ D^{*-}\pi^+$ and $D^0D^{*-}$ invariant mass
distributions for $Y(4260)\to  D^0 D^{*-}\pi^+$.  The brown dotted, green dashed and red solid lines are the 
contributions from the $S$-wave, the $D$-wave and the sum of them, respectively.}
\label{fig:InvariantMassWithSwave}
\end{figure*}

\begin{figure}[]
\centering
\vspace{0cm}
\includegraphics[width=.5\linewidth]{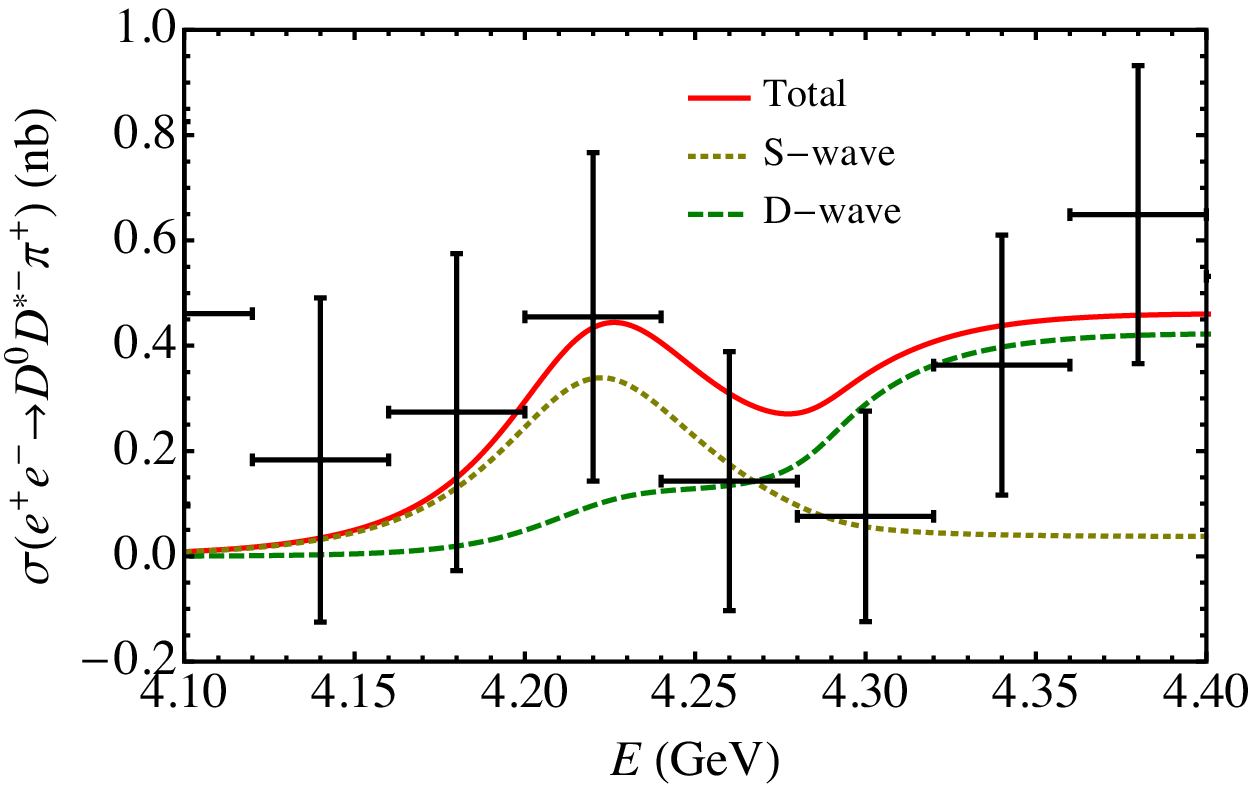}
\vspace{0cm}
\caption{Predictions for the cross section of $e^+e^-\to
D\bar D^*\pi$ with the 
dashed, dotted and solid lines denoting the contributions from $D$-wave, $S$-wave and the sum of them, respectively. Here the $D$-wave contribution is from our model due to the $D$-wave behaviour between $\bar D^*$ and $\pi$. Meanwhile the $S$-wave contribution means 
the background contribution with the $\bar D^*$ and $\pi$ in $S$-wave.
The data are from Belle~\cite{Pakhlova:2009jv}. }
\label{fig:InvariantMass}
\end{figure}
\subsection{Line shapes of the $\y$ in the $D\bar{D}^*\pi$-channel and the invariant mass distributions}
As discussed in Sec.~\ref{sec:introduction}, there are two kinds of diagrams contributing to $D\bar D^*\pi$ channels, i.e. tree diagram  (Fig.~\ref{fig:DecayDiagrams} (a)) and the $\z$ resonant contribution (Fig.~\ref{fig:DecayDiagrams} (b)). Since $D_1$ decays to $D^*\pi$ is in $D$-wave,  the contribution from the $D_1$ pole results in an enhancement
of the $D^*\pi$ spectrum at the higher mass end, i.e. approaching the $D_1$ pole
at 2.42~GeV ---  $cf$. the green dashed line in Fig.~\ref{fig:InvariantMassWithSwave} (b). Furthermore, due to the same reason, 
the lower ends of both the $D\pi$ and $D^*\pi$ invariant mass
distributions are strongly suppressed --- $cf$.  the green dashed lines in Fig.~\ref{fig:InvariantMassWithSwave} (a) and (b).

 A significant fraction of the enhancement near the $ D\bar D^*$ threshold, 
as shown by the green dashed line 
in Fig.~\ref{fig:InvariantMassWithSwave}~(c), may be understood as a reflection of the
enhancement predicted in $\bar D^*\pi$.
 With the current data quality and the level of sophistication of our model
we are not able to disentangle this from the contribution of $\z$.
The apparent discrepancy between our model prediction and the BESIII
data~\cite{Ablikim:2013xfr} should come from some $\bar D^*\pi$ $S$-wave contribution that we here include as 
 an additional, small contribution to the $\y$ wave function.
 To investigate this idea further we include this additional $S$--wave. Now  the full amplitude can be expressed as 
\begin{eqnarray}
\mathcal{M}_{D\bar{D}^*\pi}&=&\epsilon_Y^a\epsilon_{\bar D^*}^b\left [C_S\delta^{ab}+C_D(E,M_{D\bar D^*},M_{\bar D^*\pi})\left(\hat q^a\hat q^b-\frac 13 \delta^{ab}\right)\right]
\label{eq:DplusS}
\end{eqnarray}
with $C_S$ the $S$-wave strength and $C_D(E,M_{D\bar D^*},M_{\bar D^*\pi})$ the $D$-wave strength.
 The $S$-wave strength can be parameterized as 
\begin{eqnarray}
C_S=\alpha  (M_{D\bar D^*}^2+\beta)G_Z(E)
\end{eqnarray}
which respects Watson theorem as long as $\alpha$ and $\beta$ are real numbers~\footnote{The equation is adapted from Eq.~(7) of Ref.~\cite{myFF}}.   
The $D$-wave strength is extracted from our model, i.e. the sum of Fig.~\ref{fig:DecayDiagrams} (a) and (b). 

From the fit to the $D^0D^{*-}$ spectrum from $3.88~\mathrm{GeV}$ to $4.1~\mathrm{GeV}$, $cf$. Fig.~\ref{fig:InvariantMassWithSwave} (c), we obtain $|\alpha|=(6.72\pm 0.17)~\mathrm{GeV}^{-1}$, $\beta=(-15.28\pm 0.01)~\mathrm{GeV}^2$ and $\chi^2/d.o.f.=59.02/(57-3)$. 
Since in the invariant mass spectrum there is no interference between the $S$-- and the $D$--wave contribution, 
the fit does not allow one to fix the sign of $\alpha$. In what follows we chose $\alpha<0$ in order to get the correct angular distributions as discussed in the next section.
The result of the fit as well as the impact of the $S$--wave on the other invariant mass spectra 
is shown as the red solid line in the three panels of Fig.~\ref{fig:InvariantMassWithSwave}. The individual contribution from the $S$-wave is displayed by the brown dotted line. As one can see, although the small admixture of additional $S$-wave component in the $\y$ wave function will make the suppression of the lower end of the $D\pi$ and $D^*\pi$ invariant mass distribution not as significant as before, 
in the region that matters the most for both the $\y$ and the $\z$ the $D$--wave contribution still dominates, i.e. the $D_1\bar D$--component is still the most important part of the $\y$ wave function.

With all the parameters from the fits above, i.e. the line shapes of $\y$ in $e^+e^-\to Y(4260)\to J/\psi\pi\pi$ and $h_c\pi\pi$ 
processes and the invariant mass distributions in $e^+e^-\to Y(4260)\to D\bar{D}^*\pi$ process,  measured close to the pole of the $\y$,
 one can  predict also  the line shape for $D\bar{D}^*\pi$ process.
The green dashed line in Fig.~\ref{fig:InvariantMass} shows this prediction when only the $D_1\bar D$ component is included. 
 Again, a nontrivial structure arises from
the presence of the $S$-wave $D_1\bar{D}$ threshold.  Especially,  the rate predicted above the nominal $D_1\bar{D}$ threshold
is higher than at the actual $\y$ peak position.
Although this picture is changed quantitatively by the inclusion of the additional $S$--wave, the basic features remain.
Especially,  if the $\y$ is a hadronic molecule, one will not find a
Breit-Wigner line shape around 4.26~GeV in  $e^+e^-\to Y(4260)\to D\bar{D}^*\pi$.
Our prediction is consistent with the existing data ~\cite{Pakhlova:2009jv},  although improved measurements are clearly needed
to confirm or disprove our predictions.

\subsection{Angular distributions in  the $D\bar{D}^*\pi$ channel}

\begin{figure*}
\centering \includegraphics[width=0.5\linewidth]{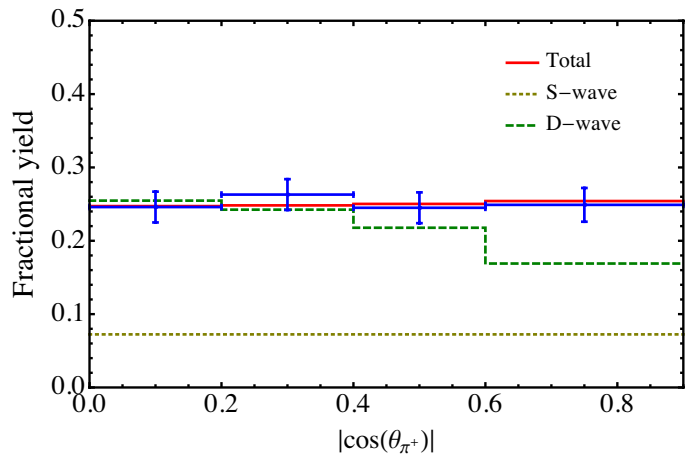}
\caption{ Angular distribution of the pion in the rest frame of the $\y$ with respect
to the beam axis (Jackson angle).
The legends are the same as that in Fig.~\ref{fig:InvariantMassWithSwave}.
The experimental data are from Ref.~\cite{Ablikim:2013xfr}. }
\label{fig:OurBachelorPion}
\end{figure*}

\begin{figure*}
\centering \includegraphics[width=1.\linewidth]{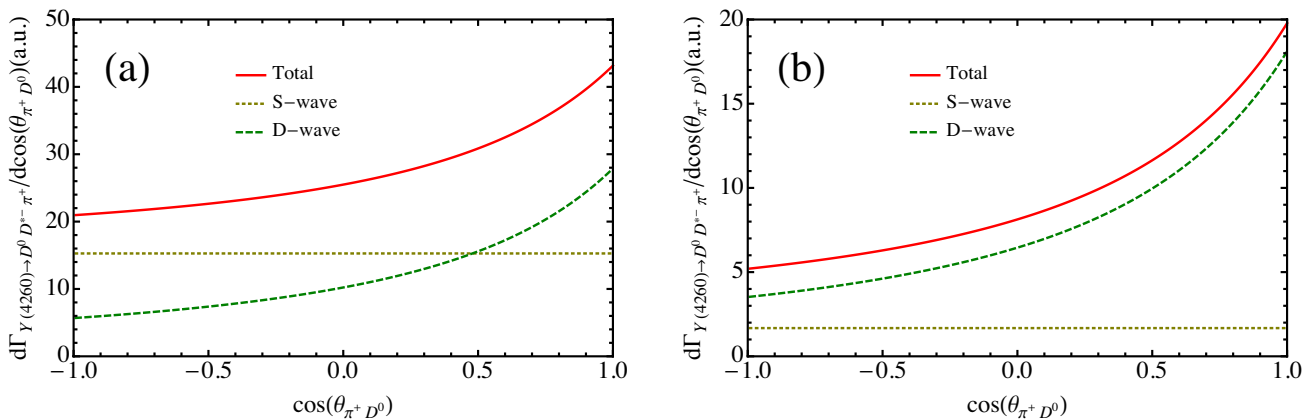}
\caption{The partial width of $\y\to D^0D^{*-}\pi^+$ is shown in terms
of the relative angle between the $\pi^+$ and the $D^0$ in the $D^0D^{*-}$ rest frame. The legends are the same as that in Fig.~\ref{fig:InvariantMassWithSwave}.
Left panel: using the whole Dalitz plot; right panel: with a cut on small $D\bar D^*$ invariant masses imposed, i.e. using the events from the $D\bar D^*$ threshold to $3.92~\mathrm{GeV}$.}
\label{fig:ThetaPi}
\end{figure*}

With the parameters fixed by the fit to the $D\bar D^*$ invariant mass distribution in the previous section, 
we now investigate  two angular distributions: the Jackson angle of the spectator pion, $\theta_\pi$, and the $D$-$\pi$ helicity angle, $\theta_{\pi D}$.

The angular distribution of the Jackson angle of the spectator pion, $\theta_{\pi}$, defined as the angle between pion and the beam direction in the overall center-of-mass frame~\cite{Ablikim:2013xfr},
is shown in Fig.~\ref{fig:OurBachelorPion}.  Since the experimental data are taken around the $\z$ pole, we integrate  the $D\bar D^*$ invariant mass from the threshold to $3.92~\mathrm{GeV}$, $cf$. Fig.~\ref{fig:InvariantMassWithSwave} (c).
 As shown in Fig.~\ref{fig:OurBachelorPion}, the large $D$-wave interfering with the strength of the  small $S$-wave fixed before
 and having chosen the sign of $\alpha$ such that there is a destructive interference between $S$-- and $D$--waves,
  leads to an almost-flat angular distribution
 (a more general discussion about the angular distribution of the spectator pion can be found in App.~\ref{app:AngularDistribution}). Note, the information  
 encoded in the Jackson angle goes beyond that contained in the Daliz plot. Thus, the agreement we find between our calculation and the
 data for $\theta_\pi$ is non-trivial, although we fit to the $D\bar D^*$--invariant mass distribution. 

The helicity angle $\theta_{\pi D}$  is defined as the relative angle between $\pi$ and $D$ in $D\bar D^*$ rest frame.
Our results for this observable are shown as the green dashed, brown dotted and red solid lines for the
$D$--wave from the molecular component of the $\y$, the additional $S$--wave and the sum of both 
are shown in Fig.~\ref{fig:ThetaPi}. 
In the left panel the whole range of invariant masses allowed kinematically is integrated. 
As one can see, the $D$--wave coming the molecular component of the $\y$ leads to a clearly
visible forward--backward asymmetry that stays prominent also after inclusion of the additional $S$ wave.
This signal can be further enhanced by imposing a cut on small $D\bar D^*$ invariant masses, i.e. from the $D\bar D^*$ threshold to $3.92~\mathrm{GeV}$, as shown in the right
panel. Clearly this observable is very sensitive to the $D_1\bar D$ component of $\y$.
So far an empirical value is available only for the
ratio~\cite{Ablikim:2013xfr}
\begin{equation}
\mathcal{A}=\frac{n_{>0.5}-n_{<0.5}}{n_{>0.5}+n_{<0.5}}=(0.12\pm0.06)
\end{equation}
 reflecting the asymmetry of  events between $|\cos (\theta_{\pi D})|>0.5$ and $|\cos (\theta_{\pi D})|<0.5$,
 where the full range of invariant masses allowed kinematically was included.
 From integrating the observable shown in the left panel of Fig.~\ref{fig:ThetaPi} we find
 $\mathcal{A}=0.0,~0.11,~0.05$ for the $S$ wave, $D$ wave and the sum of both, respectively. We can see that 
 a pure $D$-wave contribution with $\mathcal{A}=0.11$ agrees with the experimental data  perfectly.
 Nevertheless the other two results deviate by less than two sigma. Given the distributions shown in Fig.~\ref{fig:ThetaPi},
 even with the present experimental accuracy, a more decisive observable might be the forward--backward asymmetry
 of $\theta_{\pi D}$
 \begin{equation}
\mathcal{A}_{fb}=\frac{n_{>0}-n_{<0}}{n_{>0}+n_{<0}} \ .
\end{equation}
From our calculation we find 
$\mathcal{A}_{fb}=
0.0, 0.37, 0.16$ for S-wave, D-wave and the sum of them, when the whole kinematic region is integrated.

To unambiguously determine how large the $S$-wave contribution is improved data is needed. In addition, we  need to do an overall fit to all the 
available data in $J/\psi\pi\pi$, $h_c\pi\pi$, $D\bar D^*\pi$ and $D^*\bar D^*\pi$ processes which is beyond the purpose of this work.

\section{summary}

In summary, we have demonstrated in this study that, if the 
$Y(4260)$ is a $D_1\bar D$ molecule, quite unusual line
shapes  should emerge naturally in both the $J/\psi \pi \pi$ and $h_c \pi \pi$
channels. As a consequence the fits return  a pole location of the $\y$ significantly lower than 
that found in earlier studies.
In addition, the $D\bar D^*\pi$ channel is predicted to be the dominant decay mode of the $Y(4260)$.
We find that, since the $D_1\bar D$ threshold is  only a few tens of MeV
above the location of the $Y(4260)$, the $D\bar D^*\pi$ rate
 gets strongly enhanced above the nominal $D_1\bar D$
threshold. 
A detailed study of the $D\bar D^*$ invariant mass distribution revealed that
in addition to the dominant  $D_1\bar D$--component of the $\y$ that leads to
$D$--wave pions in the $D\bar D^*\pi$ channel, a subleading contribution
that produces $S$--wave pions is needed. It is important to stress that
once this additional term is fixed from the invariant mass distribution
its interference with the dominant $D$--wave term at the same time gives
a flat angular distribution for $\theta_\pi$ --- the Jacobi angle of the pion --- consistent with the data.
Fortunately there is another angular distribution, where the $D$--wave still
leads to a visible imprint, namely, the $\pi D$-helicity angle. Even when the 
$S$ wave is added, there is still a visible forward--backward asymmetry
in the prediction, which can be enhanced further by introducing a cut in
the $D\bar D^*$ invariant mass ($cf$. Fig.~\ref{fig:ThetaPi}). 
 Thus, a coherent analysis of all 
decay channels
of the $Y(4260)$ with improved data will allow one to test, if this state indeed shows a
(predominant) $D_1\bar D$--molecular structure.

\medskip

\section*{Acknowledgments}

Useful discussions with C.Z. Yuan are acknowledged.  We also acknowledge Xiao-Gang Wu for cross-checking some of the results. This work is supported, in
part, by the National Natural Science Foundation of China (Grant Nos.
11035006 and 11121092), the Chinese Academy of Sciences
(KJCX3-SYW-N2), the Ministry of Science and Technology of China
(2009CB825200), DFG and NSFC funds to the Sino-German CRC 110 ``Symmetries
and the Emergence of Structure in QCD'', and the EU
I3HP ``Study of Strongly Interacting Matter'' under the Seventh Framework
Program of the EU.

\medskip

\begin{appendix}
\section{General discussion of the $\cos \theta_\pi$ distribution}\label{app:AngularDistribution}
In this appendix a  general discussions is presented for the distribution of the 
pion angle  $\theta_{\pi}$, defined relative to the beam direction in the $e^+e^-$ rest frame for the process $e^+e^-\to \y\to\z\pi$.
  For the general amplitude of $\y\to\z\pi$ we write
\begin{eqnarray}
\mathcal{M}&=&\epsilon_Y^a\epsilon_{Z_c}^b\left(C_S\delta^{ab}+C_D\left(\hat q^a\hat q^b-\frac 13 \delta^{ab}\right)\right) \ .
\label{general}
\end{eqnarray}
The parameters $C_S$ for the $S$-wave strength and $C_D$ for the $D$-wave strength contain all information on the dynamics.
One finds
\begin{eqnarray}
\sum_\text{polarizations}|\mathcal{M}|^2=2 C_S^2-2 C_S C_D \cos^2\theta_\pi+\frac{2 C_S C_D}{3}-\frac{C_D^2
   \cos^2 \theta_\pi}{3}+\frac{5 C_D^2}{9}
\label{eq:amps}   
\end{eqnarray} 
where the sum of the polarizations 
\begin{eqnarray}
\sum_{\lambda=1,2}\epsilon_{Y}^{\lambda a}\epsilon_Y^{*\lambda b}&=&\delta^{ab}-\delta^{a3}\delta^{b3},~\quad \sum_{\lambda=1,2,3}\epsilon_{Z_c}^{\lambda a}\epsilon_{Z_c}^{*\lambda b}=\delta^{ab} \
\end{eqnarray}
were used
with the third component pointing to the beam direction. The first expression contains the fact that 
in $e^+e^-$ collisions the photon and correspondingly the $\y$ are produced transversely.
 From Eq.~(\ref{eq:amps}) one obtains a flat angular distribution not only when $C_D=0$, corresponding to a pure $S$--wave,
 but also for $C_D=-6 C_S$, where the $D$--wave dominates. 

\begin{figure}[tb]
\centering
\vspace{0cm}
\includegraphics[width=0.4\linewidth]{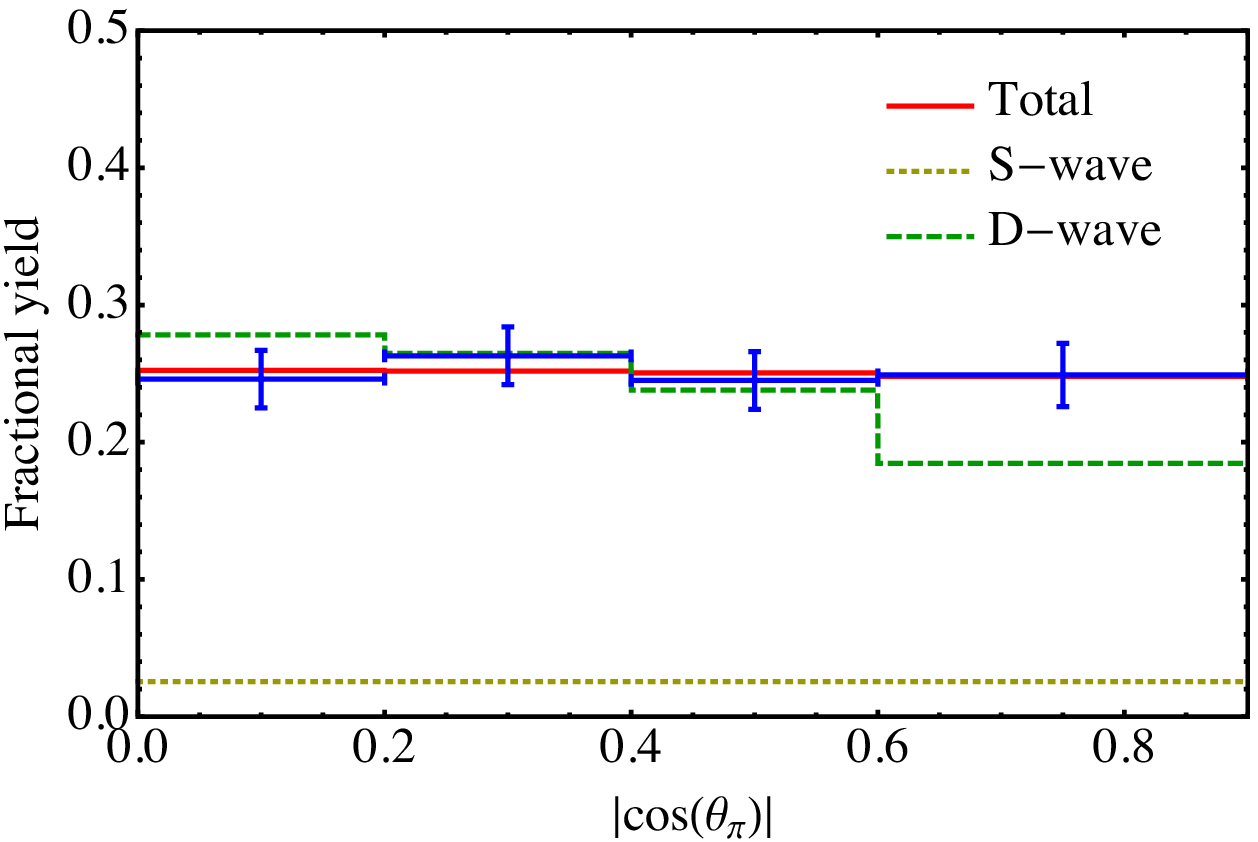} \hspace{0.1cm} \includegraphics[width=0.4\linewidth]{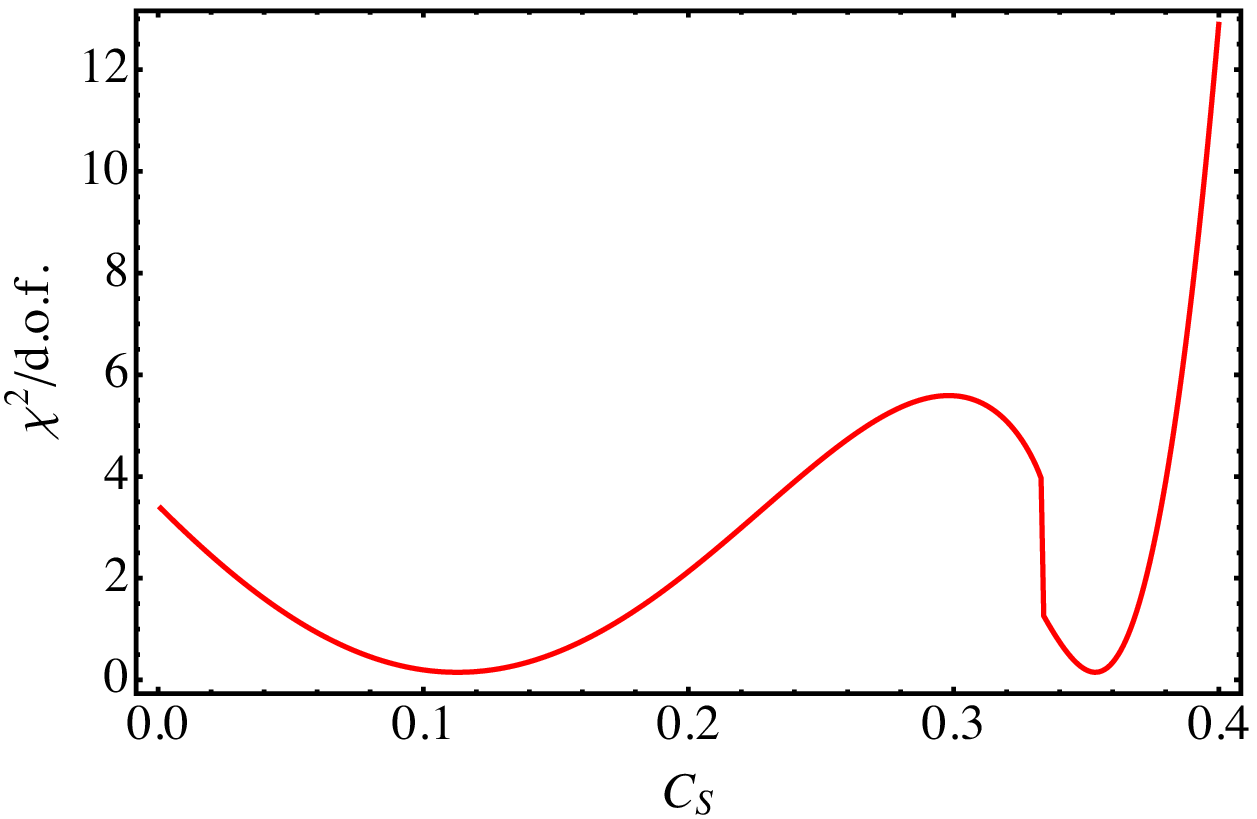}
\vspace{0cm}
\caption{Left panel: Pion angular distribution from the general amplitude of Eq.~(\ref{general}).
The green dashed, brown dotted and  red solid lines are the contributions from $D$-wave, $S$-wave and the sum of them. 
The experimental data are from Ref.~\cite{Ablikim:2013xfr}. Right panel: $\chi^2$ in terms of the $S$-wave strength parameter $C_S$ --- for
each value of $C_S$ the parameter $C_D$ is refitted. }
\label{fig:Chisquare}
\end{figure}

To be specific, we apply this general parameterization  to the data for the pion angular distribution given in Ref.~\cite{Ablikim:2013xfr}. 
A fit for the parameters $C_S$ and $C_D$ gives two solutions, $cf$. Table~\ref{tab:parameters}, in agreement with the
discussion above. The individual contributions for the best fit with $D$--wave dominance are shown in  the left panel of Fig.~\ref{fig:Chisquare}.
In the right panel of this figure we show the variation of the $\chi^2$--value, when for a fixed value of $C_S$ the parameter $C_D$ is fitted to the data. As one can see the
$\chi^2$--minimum referring to the $D$--wave dominance is rather flat --- this means that no fine tuning in the relative strength of the amplitudes
is necessary to get a flat angular distribution in
$\theta_\pi$; all it takes is an admixture of a small, non--vanishing $S$--wave amplitude.

\begin{table}[htdp]
\caption{The parameters $C_S$ and $C_D$ of the two solutions, i.e. the $S$-wave dominant and the $D$-wave dominant schemes, are listed in this table.}
\begin{center}
\begin{tabular}{|c|c|c|}
\hline\hline
parameters& $S$-wave dominant & $D$-wave dominant\\\hline
$C_S$& $ 0.35\pm 0.01$&$ 0.11\pm 0.02$ \\\hline
$C_D$ & $0.01\pm 0.04$ & $ -0.71\pm 0.01$\\\hline
$\chi^2/d.o.f.$ & 0.23 & 0.23\\ 
\hline\hline
\end{tabular}
\end{center}
\label{tab:parameters}
\end{table}%

\end{appendix}


\begin{thebibliography}{99}

\bibitem{Aubert:2005rm}
  B.~Aubert {\it et al.}  [BaBar Collaboration],
  Phys.\ Rev.\ Lett.\  {\bf 95}, 142001 (2005).

\bibitem{Brambilla:2010cs}
  N.~Brambilla {\it et al.},
  Eur.\ Phys.\ J.\ C {\bf 71}, 1534 (2011).


\bibitem{Ablikim:2013mio}
  M.~Ablikim {\it et al.}  [BESIII Collaboration],
  Phys.\ Rev.\ Lett.\  {\bf 110}, 252001 (2013).

\bibitem{Liu:2013dau}
  Z.~Q.~Liu {\it et al.}  [Belle Collaboration],
  Phys.\ Rev.\ Lett.\  {\bf 110}, 252002 (2013).


\bibitem{Xiao:2013iha}
  T.~Xiao, S.~Dobbs, A.~Tomaradze and K.~K.~Seth,
  Phys.\ Lett.\ B {\bf 727}, 366 (2013)
  [arXiv:1304.3036 [hep-ex]].
\bibitem{Ablikim:2013emm}
  M.~Ablikim {\it et al.}  [BESIII Collaboration],
  arXiv:1308.2760 [hep-ex].


\bibitem{Ablikim:2013wzq}
  M.~Ablikim {\it et al.}  [BESIII Collaboration],
  Phys.\ Rev.\ Lett.\  {\bf 111}, 242001 (2013)
  [arXiv:1309.1896 [hep-ex]].

\bibitem{Wang:2013cya}
  Q.~Wang, C.~Hanhart and Q.~Zhao,
  Phys.\ Rev.\ Lett.\  {\bf 111}, 132003 (2013).




\bibitem{Zhu:2005hp}
  S.-L.~Zhu,
  Phys.\ Lett.\ B {\bf 625}, 212 (2005).

\bibitem{Kou:2005gt}
  E.~Kou and O.~Pene,
  Phys.\ Lett.\ B {\bf 631}, 164 (2005).

\bibitem{Close:2005iz}
  F.~E.~Close and P.~R.~Page,
  Phys.\ Lett.\ B {\bf 628}, 215 (2005).



\bibitem{Voloshin:2007dx}
  M.~B.~Voloshin,
  Prog.\ Part.\ Nucl.\ Phys.\  {\bf 61}, 455 (2008).

\bibitem{Dubynskiy:2008mq}
  S.~Dubynskiy and M.~B.~Voloshin,
  Phys.\ Lett.\ B {\bf 666}, 344 (2008).

\bibitem{Li:2013ssa}
  X.~Li and M.~B.~Voloshin,
  arXiv:1309.1681 [hep-ph].

\bibitem{Ding:2008gr}
  G.-J.~Ding,
  Phys.\ Rev.\ D {\bf 79}, 014001 (2009).

\bibitem{Li:2013bca}
  M.-T.~Li, W.-L.~Wang, Y.-B.~Dong and Z.-Y.~Zhang,
  arXiv:1303.4140 [nucl-th].

\bibitem{Dai:2012pb}
  L.~Y.~Dai, M.~Shi, G.-Y.~Tang and H.~Q.~Zheng,
  arXiv:1206.6911 [hep-ph].

\bibitem{LlanesEstrada:2005hz}
  F.~J.~Llanes-Estrada,
  Phys.\ Rev.\ D {\bf 72}, 031503 (2005).

\bibitem{MartinezTorres:2009xb} 
  A.~Martinez Torres  {\it et al.},
  Phys.\ Rev.\ D {\bf 80}, 094012 (2009)
  [arXiv:0906.5333 [nucl-th]].

\bibitem{Beringer:1900zz}
  J.~Beringer {\it et al.}  [Particle Data Group Collaboration],
  Phys.\ Rev.\ D {\bf 86}, 010001 (2012).

\bibitem{Li:2013yka}
  X.~Li and M.~B.~Voloshin,
  Phys.\ Rev.\ D {\bf 88}, 034012 (2013).

\bibitem{Wang:2013kra}
  Q.~Wang {\it et al.},
  Phys.\ Rev.\ D {\bf 89}, 034001 (2014)
  [arXiv:1309.4303 [hep-ph]].

  
\bibitem{Guo:2013sya}
  F.-K.~Guo {\it et al.},
  Phys.\ Rev.\ D {\bf 88}, 054007 (2013).

\bibitem{Guo:2013zbw}
  F.-K.~Guo {\it et al.},
  Phys.\  Lett.\  B {\bf 725}, 127 (2013).

\bibitem{Yuan:2013lma}
  C.-Z.~Yuan,
  arXiv:1310.0280 [hep-ex].


  
\bibitem{Guo:2009wr}
  F.-K.~Guo, C.~Hanhart and U.-G.~Mei{\ss}ner,
  Phys.\ Rev.\ Lett.\  {\bf 103}, 082003 (2009)
  [Erratum-ibid.\  {\bf 104}, 109901 (2010)].

\bibitem{Guo:2010ak}
  F.-K.~Guo {\it et al.},
  Phys.\ Rev.\ D {\bf 83}, 034013 (2011).

\bibitem{Weinberg:1963}
 S.~Weinberg,
  Phys.\ Rev.\  {\bf 130} (1963) 776.


\bibitem{Baru:2003}
V.~Baru $et$ $al.$, 
  Phys.\ Lett.\ B {\bf 586} (2004) 53
  [hep-ph/0308129].


\bibitem{Wu:2013onz} 
  X.~-G.~Wu, C.~Hanhart, Q.~Wang and Q.~Zhao,
  Phys.\ Rev.\ D {\bf 89}, 054038 (2014)
  [arXiv:1312.5621 [hep-ph]].

\bibitem{mehen}
 T.~Mehen and J.~Powell,
  Phys.\ Rev.\ D {\bf 88} (2013) 3,  034017
  [arXiv:1306.5459 [hep-ph]].


\bibitem{Cleven:2013sq}
  M.~Cleven {\it et al.},
  Phys.\ Rev.\ D {\bf 87}, 074006 (2013).



\bibitem{Colangelo:2005gb}
  P.~Colangelo, F.~De Fazio and R.~Ferrandes,
  Phys.\ Lett.\ B {\bf 634}, 235 (2006).

\bibitem{Cleven:2011gp}
  M.~Cleven, F.-K.~Guo, C.~Hanhart and U.-G.~Mei{\ss}ner,
  Eur.\ Phys.\ J.\ A {\bf 47}, 120 (2011).


\bibitem{Yuan:2007sj}
  C.~Z.~Yuan {\it et al.}  [Belle Collaboration],
  Phys.\ Rev.\ Lett.\  {\bf 99}, 182004 (2007).

\bibitem{Aubert:2006ge}
  B.~Aubert {\it et al.}  [BaBar Collaboration],
  Phys.\ Rev.\ Lett.\  {\bf 98}, 212001 (2007).
  
   \bibitem{myFF}
   C.~Hanhart,
  Phys.\ Lett.\ B {\bf 715} (2012) 170
  [arXiv:1203.6839 [hep-ph]].
  
\bibitem{Ablikim:2013xfr}
  M.~Ablikim {\it et al.}  [BESIII Collaboration],
  Phys.\ Rev.\ Lett.\  {\bf 112}, 022001 (2014)
  [arXiv:1310.1163 [hep-ex]].


\bibitem{Pakhlova:2009jv}
  G.~Pakhlova {\it et al.}  [Belle Collaboration],
  Phys.\ Rev.\ D {\bf 80}, 091101 (2009).
  
\end{thebibliography}
\end{document}